\documentclass[12pt,preprint]{aastex}
\usepackage{natbib,psfig}

\newcommand{\agn}{{\small AGN}}
\newcommand{\uta}{{\small UTA}}
\newcommand{\hullac}{{\small HULLAC}}
\newcommand{\xstar}{{\small XSTAR}}

\newcommand{\xmm}{{\it XMM-Newton}}
\newcommand{\rgs}{{\small RGS}}
\newcommand{\chandra}{{\it Chandra}}
\newcommand{\letg}{{\small LETG}}

\begin{document}

\title{Soft X-Ray Absorption by Fe$^{0+}$ to Fe$^{15+}$ in Active Galactic Nuclei}
\author{Ehud Behar \altaffilmark{1}, Masao Sako \altaffilmark{1}, Steven M. Kahn \altaffilmark{1}}

\altaffiltext{1}{Columbia Astrophysics Laboratory and Department of Physics,
                 Columbia University, 550 West 120th Street, New York, NY
                 10027; behar@astro.columbia.edu (EB), 
	        masao@astro.columbia.edu (MS), skahn@astro.columbia.edu (SMK)}

\received{}
\revised{7/4/01}
\accepted{}

\shorttitle{Soft X-ray Absorption by Fe}
\shortauthors{Behar et al.}

\begin{abstract}

  A full set of calculations is presented for inner-shell $n$~= 2 to 3
  photoexcitation of the 16 iron charge states: Fe$^{0+}$ (Fe {\small I}) through
  Fe$^{15+}$ (Fe {\small XVI}). The blend of the numerous absorption lines arising from
  these excitations (mainly 2p~- 3d) forms an unresolved transition array (\uta), which has been recently identified as a prominent feature between 16~- 17~\AA\ in the soft X-ray spectra of active galactic nuclei (\agn). Despite the blending within charge-states, the ample separation between the individual-ion features enables precise diagnostics of the ionization range in the absorbing medium. Column density and turbulent velocity diagnostics are also possible, albeit to a lesser accuracy. An abbreviated set of atomic parameters useful for modeling the Fe 2p~- 3d \uta\ is given. It is shown that the effects of accompanying photoexcitation to higher levels ($n~\ge$~4), as well as the associated photoionization edges, may also be relevant to AGN spectra.

\end{abstract}


\section{INTRODUCTION}
\label{sec:intro}

Active galactic nuclei (\agn) viewed directly towards the supermassive black
hole (e.g. Seyfert~1 galaxies) can produce a rich absorption spectrum. High-flux X-ray and UV continuum emission from the central source ionizes the ambient gas to high charge states. Consequently, the ionized gas along the line of sight imposes absorption features on the observed continuum. Away from the central source, the ionizing flux gradually diminishes, limiting the maximum charge states that can be formed, especially for high-Z elements. Notwithstanding, the photon energy distribution, generally a power law, allows for a substantial fraction of energetic photons. Therefore, the intermediately ionized ions are exposed to X-ray photons that can be appreciably more energetic than the typical energies for ionization or excitation from the valence shells of these ions. When the column density through this gas is sufficient, absorption by means of inner-shell photoexcitation processes plays an important role in the formation of discrete features in the spectrum.

In particular, the $n$~= 2 to 3 absorption lines ($n$ being the principal
quantum number of the active electron) of the cosmically abundant element iron
(Fe) fall in the soft X-ray band that can be readily measured by the grating spectrometers on board \chandra\ and \xmm. The $n$~= 2 to 3 valence
transitions are well known in the soft X-ray band for the (L-shell) ions
Fe$^{16+}$ to Fe$^{23+}$. The analogous, but inner-shell, excitations in
Fe$^{0+}$ to Fe$^{15+}$ (referred to hereafter in shorthand as Fe M-shell
ions), have not been given the same attention. In the laboratory, \citet{cp01} studied
the opacity of a thin, laser heated Fe foil illuminated by a strong X-ray
backlighter. They detected unresolved inner-shell $n$~= 2 to 3 photoabsorption
by Fe$^{4+}$ to Fe$^{9+}$ around 17~\AA\ and modeled it assuming local
thermodynamic equilibrium. Recently, we have reported the first
astrophysical detection of a similar feature in the spectrum of the luminous
quasar IRAS 13349+2438 \citep{sako01} obtained with the Reflection Grating
Spectrometer (RGS) on board \xmm. In that spectrum, we ascribe the broad
absorption trough between 16~- 17~\AA\ to an unresolved transition array (\uta)
of inner-shell $n$~= 2 to 3 (mainly 2p~- 3d) lines pertaining to various Fe
ions. As shown in \citet{sako01}, the exact shape of this absorption
feature is well fitted with a superposition of lines from Fe$^{6+}$ to
Fe$^{11+}$ with ionic column densities of 0.9~- 6.4 $\times~10^{16}$~cm$^{-2}$, which correspond roughly
to column densities of hydrogen atoms of a few times 10$^{21}$~cm$^{-2}$. Recently, the
soft X-ray spectrum of the Seyfert~1 NGC 3783 obtained with the RGS has
revealed a similar, but even more prominent Fe 2p~- 3d UTA \citep{bl01}. This feature, unaccounted for in the analysis of the \chandra\
spectrum of NGC 3783 \citep{kaspi00,kaspi01}, could explain the discrepancies
between the data and the model around 17~\AA\ in that work. A much weaker Fe 2p~- 3d UTA was tentatively identified in the Seyfert~1 Markarian 509 by \citet{pounds01}. Fig.~1 shows the Fe 2p~- 3d UTA region of the \rgs\ spectrum for both IRAS 13349+2438 and NGC 3783. Despite the apparent difference in absolute flux, the shape and structure of the UTA are somewhat similar, covering approximately the same range of ionization. The transmission in NGC 3783 is evidently lower. 

For both IRAS 13349+2438 and NGC 3783, as well as for other \agn, the Fe 2p
- 3d \uta\ can be used to diagnose the ionization state and column density for
the relatively low-ionization components of the soft X-ray absorbing
gas. Since the Fe M-shell ions also absorb in the UV and EUV bands, they can
be used to relate the UV absorber with the X-ray absorber, which have
curiously shown different column densities for many sources. Additionally, the
same charge states that produce the Fe 2p - 3d \uta\ absorption also produce
strong emission lines in the various optical bands. If both features are
observed simultaneously, the Fe 2p - 3d \uta\ can potentially provide
an independent probe of the level of ionization in the optically emitting
gas. Furthermore, if the low-ionization Fe M-shell gas is associated with the ionized skin
of the putative \agn\ torus, as postulated by \citet{sako01}, it can provide a
constraint on the size of the torus and the inclination angle of the (host)
galactic plane with respect to our line of sight.

Photoabsorption from the L-shell ($n$= 2) of neutral (or nearly neutral) Fe
falls between 17 and 18~\AA. This was most recently detected by
\citet{paerels01} in the low-mass X-ray binary X0614+091. \citet{paerels01} mention only the L-shell photoionization edges, probably because of the lack
of atomic data for the associated (photoexcitation) lines. Owing to their much
higher cross sections, lines should always appear when edges are
detected. Thus, in addition to \agn\ observations, the present calculations are
also relevant to spectra produced by absorption of local X-ray sources, such
as X0614+091, by interstellar Fe in our galaxy. This paper provides an account
of the atomic processes and the full details of the spectral modeling issues
related to the inner-shell $n$~= 2 to 3 absorption lines of Fe M-shell ions. A compact set of wavelengths and oscillator strengths is made available for fitting this feature to observed spectra and for diagnostics. 

\section{ATOMIC CALCULATIONS}
\label{sec:atomic_calc}

In order to calculate the wavelengths and oscillator strengths for the
radiative transitions, we use the multi-configuration, relativistic \hullac\
(Hebrew University Lawrence Livermore Atomic Code) computer package \citep
{bs01}. The intermediate-coupling level energies are calculated using the
relativistic version \citep{koenig72, klapisch77} of the Parametric Potential
method by \citet{klapisch71}. The relativistic approach employed in \hullac, which uses the jj coupling scheme for its initial vector basis, is perfectly adequate for the inner-shell excitations studied in this work. The high efficiency of \hullac\ is also crucial, owing to the numerous transitions and the considerable complexity of the electronic structure in the ions considered. Calculations are carried out for neutral iron (Fe$^{0+}$) through Na-like Fe$^{15+}$. The lower Fe charge states could be less relevant to the \agn\ vicinity, but more to the narrow line regions or to X-ray illuminated interstellar gas.

The atomic structure calculations include the ground configurations
2s$^2$2p$^6$3l$^x$ ($x$~= 1 to 14) for Fe$^{15+}$ through Fe$^{2+}$, and
2s$^2$2p$^6$3l$^{14}$4s$^x$ ($x$~= 1, 2) for Fe$^{1+}$ and Fe$^{0+}$. For the
ground configurations of Fe$^{1+}$ and Fe$^{0+}$, the [3d + 4s] configuration
mixings are included, i.e., 2s$^2$2p$^5$3s$^2$3p$^6$[3d$^6$4s$^x$ +
3d$^7$4s$^{x-1}$] ($x$~= 1, 2). All of the $n$~= 2 to 3 photoexcitation
transitions from the ground level are calculated for each ion, among which the
2p-3d and 2s-3p are the strongest. For charge states Fe$^{8+}$ and lower, the
3p subshell is full and, thus, the 2s-3p excitation channel is not
possible. For Fe$^{15+}$, the 2p-3s excitation channel is open and is also
calculated. For the 2p-excited Fe$^{10+}$~- Fe$^{15+}$ ions, the important
[3s3d + 3p$^2$] configuration mixings are possible and are included, i.e.,
2s$^2$2p$^5$[3s$^2$3p$^x$3d + 3s3p$^{x+2}$] ($x$~= 0 to 4), as well as
2s$^2$2p$^5$[3s3d + 3p$^2$] for Fe$^{15+}$. The significance of other mixing
effects, such as [3p$^2$ + 3d$^2$], which were found important for excitations
out of the 3p subshell \citep {quinet95}, were also examined. We find
these mixings to reduce the total $n$~= 2 to 3 photoexcitation effect by about
1\% and to alter the level energies by much less than that, while increasing
the number of absorption lines by a factor of ten and more. Therefore, for the
sake of simplifying the computations, we have excluded these mixings. Electric
and magnetic dipole and quadrupole transitions were computed. The
contribution, however, of excitations other than electric-dipole is less than
0.1 \%.

The inner-shell excited levels studied here tend to strongly autoionize. This
has several effects on the observed spectrum; First, through the low
fluorescence yield. Since the inner-shell $n$~= 2 to 3 photoexcitation processes are by and large followed by autoionization rather than X-ray re-emission, they leave an observable imprint only in absorption, but not in emission. This absence of fluorescence following L-shell photoexcitation is evident in the soft X-ray spectra of Seyfert~1 galaxies, but even more so in the Seyfert~2 spectra, where emission is not confused with absorption. The low fluorescence yield further implies that dielectronic recombination via these channels is inefficient. Another consequence of the high autoionization rates is the broad natural widths of the absorption lines associated with inner-shell photoexcitation. At high column densities, the natural width of the line can become important in determining the total flux absorbed. For many of the transitions considered in this work, the total autoionization rate is an order of magnitude higher than the radiative decay rate. Thus, in order to obtain correct line profiles and to account for the total photoabsorption effect, we have calculated the total depletion rate incorporating both the autoionization and radiative decay rates to obtain the natural line width. The natural widths obtained here could be, in some cases, a slight underestimation of the actual width because including all of the autoionization transitions is computationally not permissible. The ionized configurations that were included in the autoionization calculations were the most important ones that can be reached by a single-electron transition from the ground level.

The results of the above-described calculations are shown in Figs. 2 and 3,
where the absorption spectra produced by the inner-shell $n$~= 2 to 3
excitations are presented for Fe$^{0+}$ through Fe$^{15+}$. The complex atomic
structure and the vast amount of detail in the spectra can be clearly seen in
the figures. For display purposes, a low ionic column density of $N_{ion}$~= 2
$\times10^{15}$~cm$^{-2}$ and a temperature of $kT_{ion}$~= 10~eV are chosen,
with no turbulent velocity. These parameter values help minimize the effects
of saturation and blending of the individual spectral lines in the plot. The
dependence of the observed spectrum on these parameters is discussed below in more detail.

The largest group of lines for each ion represents the 2p~- 3d
transition array, and can be seen in the figures to shift from 
$\sim$~17.5~\AA\ for Fe$^{0+}$ to $\sim$~15.2~\AA\ for Fe$^{15+}$. In fact, the
predominant transitions are 2p$_{3/2}$~- 3d$_{5/2}$ and 2p$_{1/2}$~-
3d$_{3/2}$ . The 2p$_{3/2}$~- 3d$_{3/2}$ lines are significantly
weaker, although mixing among the jj configurations makes it
difficult, occasionally, to unambiguously identify a line with a
single-electron transition. The observable isonuclear trend of the 2p
- 3d transition array, shifting to shorter wavelengths as the Fe ion
charge increases, can be used to accurately diagnose the ionization
range of the absorbing plasma. The spectral resolution of the
contemporary soft X-ray grating spectrometers on board \chandra\ and
\xmm\ are sufficient for measuring the trough of the Fe 2p~- 3d \uta\ rather precisely. Starting from Fe$^{9+}$ ($\sim$~14.9~\AA), where the 3p subshell has one vacancy, and continuing to higher charge states ($\sim$~14.0~\AA\ for Fe$^{15+}$) with more 3p vacancies, the 2s~- 3p lines, mostly 2s~- 3p$_{3/2}$, can also be seen in Fig.~3. These lines will usually not be detectable in the spectrum, especially since they overlap with stronger absorption lines that arise from 2p excitations to high-$n$ levels (see Sec.~5). Fe$^{15+}$ has also two 2p-3s lines calculated at 17.395~\AA\ and 17.095~\AA, which are not shown in Fig.~3.
	
The present atomic data were compared with results of similar, but partial,
calculations by \citet {raassen01} for several ions in the Fe$^{6+}$~-
Fe$^{14+}$ range. \citet {raassen01} used Cowan's atomic code \citep{cowan81}, which employs the LS
angular momentum coupling scheme with relativistic corrections. The mean
wavelengths for individual ions agree to within a few m\AA\ between the two
sets of results. The total oscillator strengths obtained with Cowan's code are
systematically higher than the present results. The highest discrepancy of 15
\% occurs for Fe$^{10+}$, but for the most part the discrepancies are about 5
- 10 \%. These discrepancies could be explained by the different
methods, but also by \citet {raassen01} not including the important [3s3d + 3p$^2$] configuration mixings.

\section{SPECTRAL MODELING}
\label{sec:modeling}

The optical depth $\tau(\nu)$ at frequency $\nu$ is defined through the
exponential attenuation $I(\nu) = I_0(\nu) exp[-\tau(\nu)]$, where $I_0(\nu)$ and $I(\nu)$ are the unabsorbed and attenuated intensities, respectively.

\subsection{Individual line absorption}
\label{subsec:line_abs}

The optical depth $\tau_{ij}(\nu)$ due to an absorption line ($i~\to j$) can be written as: 
\begin{equation}
\tau_{ij}(\nu) = N_{ion}\sigma_{ij}(\nu)
\end{equation}
where $N_{ion}$ is the ionic column density towards the source (in cm$^{-2}$)
and $\sigma_{ij}(\nu)$ is the cross section for photoexcitation (in cm$^2$)
from $i$ to $j$. The effect of more than one line at a given frequency is multiplicative (a product of exponents). The photoexcitation cross section is:
\begin{equation}
\sigma_{ij}(\nu) = \frac{\pi e^2}{m_ec}f_{ij}\phi(\nu)
\end{equation}
Here, $e$ and $m_e$ are the electron charge and mass, $c$ is the speed of
light, $f_{ij}$ is the line oscillator strength, and $\phi(\nu)$ is the normalized line profile, generally described by a Voigt function, which is the convolution of the (Gaussian) Doppler broadening due to temperature and turbulence and the (Lorentzian) natural width of the line. The Doppler broadening can be characterized by a Doppler width $\Delta\nu_D$:
\begin{equation}
\Delta\nu_D = \frac{1}{\lambda_{ij}}\sqrt{\frac{2kT_{ion}}{M_{ion}}+v^2_{turb}}
\end{equation}
where $\lambda_{ij}$ is the line wavelength, $T_{ion}$ and $M_{ion}$ are,
respectively, the temperature and mass of the absorbing ions, and $v_{turb}$
is the turbulent velocity. Throughout this work we use $kT_{ion}$ = 10~eV. In typical absorbing, photoionized, \agn\ material, turbulence broadening ($v_{turb}~\ge$~100~km/sec) totally dominates over temperature broadening. The natural width $\Gamma_j$ (in sec$^{-1}$) is associated with the total rate for depletion of the upper level $j$ of the transition: 
\begin{equation}
\Gamma_j = \sum_{i^{\prime} < j}A_{ji^{\prime}} + \sum_{k^{\prime} < j}A^a_{jk^{\prime}}
\end{equation}
In the present work, we include both the radiative ($A$) and autoionization ($A^a$) rates for determining the natural width.

\subsection{Ionization Parameter Zones}
\label{subsec:zones}

Due to the comparable ionization energies of the Fe M-shell ions and to the
broadband nature of the ionizing flux, in physical photoionized plasmas,
several charge states will always be present simultaneously, even if the gas
is well confined in space and uniform in density and in temperature. Hence,
the observed Fe 2p~- 3d \uta\ will necessarily be a blend of several charge
states. The spectral location and shape of this feature depend on the
ionization parameter. We adopt the definition of the ionization parameter:
$\xi = L / n_er^2$ from \citet {tarter69}, where $L$ is the
luminosity of the ionizing source, $n_e$ is the electron density, and $r$ is
the distance from the source. The photoabsorption spectra expected for $\xi$
= 1, 3, and 10~erg~sec$^{-1}$~cm are plotted in Fig.~4. The total Fe column
density is taken to be 2$\times$10$^{17}$~cm$^{-2}$ and the ionic fractional
abundances are calculated with \xstar\ \citep {kallman95}. The Fe 2p~-
3d \uta\ can be seen to shift towards shorter wavelengths as $\xi$
increases. The dominant contributions for $\xi$~= 1, 3, and 10~erg~sec$^{-1}$~cm come, respectively, from Fe$^{5+}$~- Fe$^{7+}$, Fe$^{7+}$~- Fe$^{10+}$, and
Fe$^{10+}$~- Fe$^{13+}$. For $\xi$~= 10~erg~sec$^{-1}$~cm, the weak 2s~- 3p
lines below 15~\AA, which become available for Fe$^{9+}$ and higher charge
states, can also be seen in the bottom panel of Fig.~4. The Fe 2p~- 3d \uta\
broadens with $\xi$ due to the increasingly larger spacing between charge
states for the high-ionization species (see Table~1 in Sec. 4). Despite the
various ions partially overlapping in the spectrum, the imprints of individual ions can
still be noticed, even for a turbulent velocity of 500~km/sec that was used
for Fig.~4. Each dip in the theoretical spectra in Fig.~4 can be ascribed
primarily to one charge state. For the correspondence between (average)
wavelength and charge state see Table~1 in Sec.~4. For single-$\xi$ systems
with comparable velocities ($\le$~700~km/sec), these individual ion
contributions could be just marginally resolved. For a detailed analysis of
the Fe 2p~- 3d \uta\ in NGC 3783 and how it breaks up into single-ion
contributions, see \citet {bl01}.

Interestingly, all currently known detections of the Fe 2p~- 3d \uta\ are more
or less consistent with a single ionization parameter of $\xi \cong$ 3~erg~sec$^{-1}$~cm, while lower and higher ionization states are clearly not
observed. In fact, the calculated \uta\ for $\xi$~= 3~erg~sec$^{-1}$~cm in
Fig.~4, which falls roughly between 16 and 17~\AA, resembles the observed
\uta\ in both NGC 3783 and IRAS 13349+2438 (see Fig.~1) fairly
accurately. Particularly tantalizing is the absence from the soft X-ray
spectra of the ions Fe$^{12+}$~- Fe$^{15+}$, while both lower and higher
(Fe-L) charge states show prominent absorption features. This discontinuity in
ionization could be indicative of thermal instabilities precluding these ions
from co-existing with the adjacent charge states. Investigating this possibility is
beyond the scope of this paper, but preliminary calculations show that thermal
instabilities could arise, depending very sensitively on the ionizing
spectrum. The absence of charge states lower than Fe$^{6+}$, on the other
hand, is easier to understand, as a consequence of the typical X-ray fluxes in
the \agn\ vicinity, which will instantly ionize a few electrons from the Fe
atoms exposed to it. In regions further out from the center, where $\xi$ could
be much lower, the column density through the outflowing and absorbing gas
diminishes rapidly and, therefore, it is not observed.

\subsection{Equivalent Width}
\label{subsec:ew}

It is customary to assess the total amount of absorption with the equivalent
width (EW), which is essentially the area under the (normalized) transmission
curve (e.g., Figs. 2 and 3). The EW due to inner-shell $n$~= 2 to 3
photoexcitation is plotted in Fig.~5 as a function of $N_{ion}$ (the so-called
curve of growth) for five selected Fe M-shell ions at $v_{turb}$~= 500~km/sec. The EW behavior of these ions represents the EW trend for the other Fe
M-shell ions as well. The EW is calculated in the 10~- 20~\AA\ range only,
which is adequate because at the column densities considered for Fig.~5, the
line wings do not digress much from that region. The EW can be seen to
generally increase with the charge state. This is due to the gradual increase
in the total oscillator strength with the ionic charge, which, in turn, is
mostly attributed to the opening of more and more vacancies in the $n$~= 3
shell. No increase in the oscillator strength occurs from Fe$^{0+}$ to
Fe$^{2+}$, because the ground configurations for those ions differ only by
their 4s shell, which does not affect the $n$~= 2 to 3 excitation. As seen in
Fig.~5, saturation occurs at about 10$^{17}$~cm$^{-2}$, except for Fe$^{8+}$,
which has (a closed 3p-subshell and thus) only three 2p~- 3d lines (see Fig.~3). The total oscillator strength is concentrated mostly in two of these lines. Consequently, they tend to saturate at lower column densities, as can be seen in Fig.~5. Note that at very high column densities, photoexcitation gradually plays a lesser role, since the absorption in photoionization edges (not included in Fig.~5) exceeds that in the lines by far. This effect is discussed further in Sec.~5.

\section{ABBREVIATED DATA SET}
\label{sec:short_data}

There are, generally, several dozens of inner-shell $n$~= 2 to 3 lines for each Fe
M-shell ion. Typically, 20 lines from each ion account for more than 90 \% of
the EW in typical \agn\ conditions, with the exception of Fe$^{5+}$,
Fe$^{10+}$, and Fe$^{11+}$, where the 20 strongest lines represent a somewhat
smaller fraction of that effect. As can be seen in Figs. 2 and 3, most of the
lines for each ion are concentrated around a narrow spectral region, which can
be very close to the corresponding regions for the adjacent charge
states. Since several Fe M-shell ions are always present in the plasma
simultaneously and since typical \agn\ turbulent velocities of a few
100~km/sec broaden the spectral lines, the entire ensemble of these lines
pertaining to different charge states is coalesced into a broad \uta. This is
independent of the spectral resolution of the measuring device. Consequently,
the full set of the current atomic data, including its large number of lines
for each ion, would in many cases be unnecessary. Alternatively, mean wavelengths and summed oscillator strengths could be adequate for modeling and for spectral fitting purposes.

\citet {ba85} proposed compact formulas for the
mean wavelengths and spectral widths of transition arrays. However, since we
have already calculated all of the lines level by level, we can simply use our
detailed results to obtain these quantities. Additionally, the sum rules in
\citet {ba85} do not include the effects of configuration mixing. In
order to make the atomic data manageable, we have divided the lines of each
charge state into a few groups. Each group is represented by the total
oscillator strength of lines in the group $\Sigma f_{ij}$ and by an $f$-value
averaged wavelength $\lambda_{av}$ :
\begin{equation}
\lambda_{av} = \frac{\sum \lambda_{ij}f_{ij}}{\sum f_{ij}}
\end{equation}
The sums are carried out over upper levels $j$, while the lower level $i$,
here, is always the ground level. The effective natural width of a group $\Gamma_{eff}$ is defined by:
\begin{equation}
\Gamma_{eff} = \frac{\sum \Gamma_j f_{ij}}{\sum f_{ij}}
\end{equation}
Each group also has a statistical spectral width $\Delta \lambda$ defined by its $f$-weighted standard deviation:
\begin{equation}
\Delta \lambda = \sqrt{\frac{\sum \left( \lambda_{ij} - \lambda_{av} \right) ^2 f_{ij}}{\sum f_{ij}}}
\end{equation}
Assuming that the distribution of lines within a group is roughly normal
(Gaussian), and for mere modeling purposes, the statistical width can be added
to the Doppler width $\Delta \nu_D$ as an additional "velocity" component [compare with Eq. (3)]:
\begin{equation}
\Delta\nu_D = \frac{1}{\lambda_{ij}}\sqrt{\frac{2kT_{ion}}{M_{ion}} +
v^2_{turb} + 2 \left( c \frac{\Delta \lambda}{\lambda_{av}} \right)^2}
\end{equation}
The factor 2 in front of the last term in Eq. (8) is needed to properly
translate $\Delta \lambda$ defined in Eq. (7) to the usual velocity dependence
used for $\Delta \nu_D$.

The groups of $n$~= 2 to 3 absorption lines for Fe M-shell ions in the ground
level are listed in Table~1. In the first three columns, the Fe ion is given
together with the configuration (in the jj coupling scheme) and total angular
momentum quantum number $J_G$, of the ground level. The $n$~= 1 and 2 inner
electronic shells not specified in the table are assumed to be
full. Subsequently, the average wavelength $\lambda_{av}$, the total
oscillator strength $\Sigma f_{ij}$, the effective natural width
$\Gamma_{eff}$ (due to autoionization and radiative decay), and the
statistical width $\Delta \lambda$, are given for each group. The last two
columns in Table~1 indicate the leading radiative transition in the group and
the total number of (electric dipole) lines it represents. Both for the highly
ionized and for the close-to-neutral species, the separation into groups
nicely traces the spin-orbit array splitting discussed in \citet {ba85}, where
each group can be characterized by a single jj electron transition (e.g.,
2p$_{1/2}$~- 3d$_{3/2}$ and 2p$_{3/2}$~- 3d$_{5/2}$). For intermediate charge
states Fe$^{10+}$ to Fe$^{13+}$, where the [3s3d + 3p$^2$] configuration
mixing is strong, the actual atomic levels are often composed of components
from more than one configuration, making the single-electron transition
identification not absolutely strict. For those ions, the lines do not split
as nicely into spin-orbit arrays, as can be seen in Fig.~3.

Table~1 shows that the total oscillator strength generally increases with the ionic charge, while the natural width dominated by the total autoionization rates, remains more or less constant. Except for the lowest charge states (Fe$^{0+}$~- Fe$^{4+}$), the groups of strong 2p~- 3d lines from each ion are nicely separated by more than 0.1~\AA, easily resolvable with \chandra\ and \xmm, enabling an unambiguous measurement of the range of charge states existent in the absorbing plasma. Although the accuracy of the relativistic parametric potential method in \hullac\ might slightly diminish as the charge state approaches neutral, for the present inner-shell transitions, these inaccuracies are expected to be much less than 0.1~\AA.

As mentioned above, the statistical widths given in Table~1 facilitate the use
of the present atomic data for spectral analysis. We compared the curves of
growth obtained with the full atomic data set to those obtained with the
abbreviated data set for several cases. The results agree to better than a few
percent for $N_{ion}$ up to 10$^{16}$~cm$^{-2}$ (and then again for $N_{ion}
> 10^{19}$~cm$^{-2}$). On the other hand, since the largest statistical
widths of $\Delta \lambda~\sim$ 0.1~\AA, which occur for the strong groups of
Fe$^{10+}$~- Fe$^{13+}$, correspond roughly to velocities of 2500~km/sec,
for these ions, the abbreviated atomic data set given in Table~1 will
reproduce the saturation of the curve of growth correctly only for similarly
high velocities. For lower velocities, the shortened method will slightly
overestimate the column density at which saturation takes place. In the most
extreme cases we checked:  Fe$^{10+}$~- Fe$^{13+}$ at 100~km/sec, the
shortened method can overestimate the actual EW by as much as 40 \% at a
column density of 10$^{17}$~cm$^{-2}$. For lower charge states or higher
velocities, the agreement is much better in the entire range of column
densities. In short, although the grouping of lines can be useful for spectral
fitting and modeling, the shortened method should be used with caution,
particularly for low velocity systems ($\le$~100~km/sec) and column
densities of 10$^{17}$~- 10$^{18}$~cm$^{-2}$.

\section{ADDITIONAL AND RELATED ABSORPTION}
\label{sec:high_n}

It is important to note that the present set of calculations is by no means
the complete picture for absorption by Fe M-shell ions. The $n$~= 2 to 3 transition array is, indeed, the most conspicuous in the soft X-ray spectrum, but will be regularly accompanied by weaker lines emerging from excitations to higher levels. At high column densities, the photoionization edges could also play a major role. Additionally, even for moderately dense plasmas, the population of ions in excited levels, especially metastable levels, which occasionally lie extremely close in energy to the ground level, can be sufficient to produce photoabsorption from excited levels. This would be due to collisional- and photo- excitation as well as radiative cascades towards these levels. In these cases, the absorption spectrum can be calculated only after the excited level populations are determined through comprehensive collisional-radiative models (that include photoexcitation) for each ion.

Full models for all of the ions are beyond the scope of this paper, but
we do wish to give an idea of the total photoabsorption effect from the ground
level. For this purpose, we choose Fe$^{7+}$ for its relatively simple atomic
structure. The absorption spectrum is calculated including inner-shell 2p~-
$n$d photoexcitation for $n$~= 3 through 9, as well as the photoionization
edges at the $n~\to \infty$ series limit. All these are readily obtained with
the \hullac\ code. The results are presented in Fig.~6 for three ionic column
densities of 3$\times$10$^{16}$, 10$^{17}$, and 10$^{18}$~cm$^{-2}$. The
latter column density is rarely relevant to \agn\ sources, but nonetheless is
used here to exemplify the high-column limit. The turbulent velocity used to
calculate the transmission functions in Fig.~6 is fixed at 100~km/sec. At a
column density of 3$\times$10$^{16}$~cm$^{-2}$, where the 2p~- 3d lines are
not yet saturated, the high-$n$ lines can be seen to contribute to the
absorption spectrum. In fact, the high-$n$ lines can potentially give a
sensitive indication of the ionic column density in the region around 10$^{17}$
cm$^{-2}$, where the 2p-3d \uta\ is saturated. The photoionization edges,
however, although contributing to the EW, are spread out and
therefore will be hard to detect. Note the unabsorbed region of the spectrum
around 16~\AA\ between the 2p~- 3d (\uta) and the 2p~- 4d features. In the soft
X-ray spectra of many \agn, but particularly in the spectrum of NGC 3783 where the Fe 2p~- 3d \uta\ is deep
\citep {bl01}, the absence of absorption in this region
creates a hump that could be mistakenly interpreted as emission. At very
high column densities, the photoionization edges dominate the lines as seen in
Fig.~6. Both the L-shell edges (photoionization from $n$~= 2) in the soft
X-ray band, but also the M-shell edges ($n$=~3) that occur at longer
wavelengths ($>$ 50~\AA), can have an appreciable effect on the observed spectrum between 10 and 20~\AA. Both these effects can be seen in the bottom panel of Fig.~6 in the high column density plots for $N_{ion}$~= 10$^{18}$~cm$^{-2}$. At these high column densities, even many of the high-$n$ lines are saturated.

In order to quantitatively compare the effect of the 2p~- 3d \uta\ with those
of the high-$n$ lines as well as with those of the photoionization edges, we
plot in Fig.~7 the EW as a function of ionic column density, separately for
the 2p~- 3d lines, the 2p-$n$d ($n$~= 4~- 9) lines, and the L-shell and M-shell
photoionization edges. The total EW is also shown. Similarly to Fig.~6, the
turbulent velocity is fixed at 100~km/sec. It can be seen that up to $N_{ion}$
= 10$^{17}$~cm$^{-2}$, the 2p~- 3d lines dominate the EW. At higher column
densities, when the 2p~- 3d lines saturate, both the L-shell and M-shell
photoionization edges give the predominant absorption effect. The high-$n$
lines give a rather small contribution to the EW, about half that of the 2p~-
3d lines, although judging from Fig.~6, they should be detectable at ionic column
densities of $\sim$~10$^{17}$~cm$^{-2}$. These numbers could vary slightly from ion to ion, but the general trends illustrated in Fig.~7 are representative of the EW behavior for all of the Fe M-shell ions. For ions, such as Fe$^{8+}$, in which the 2p-3d lines saturate at lower column densities, the relative contribution of the high-$n$ lines would be more significant.

Note that analogous 2p - 3d \uta\ absorption can also be produced by elements other than
Fe. However, in order for those to be observed, many M-shell ions pertaining
to those elements need to show high EW in the inner-shell 2p - 3d lines, which in turn requires a much lower ionization parameter than
that for the Fe 2p - 3d \uta. To give a general idea of where these lines
could be expected in the spectrum, we calculated the wavelengths for the strongest
inner-shell 2p - 3d lines of (Mg-like) Ca$^{8+}$, Ar$^{6+}$, S$^{4+}$, and
Si$^{2+}$. The approximate wavelengths are, respectively: 31.4 \AA, 43.5 \AA,
64.0 \AA, and 106 \AA. Of these, only the Ca 2p - 3d \uta\ falls in the \rgs\ wavelength
band, while the others could be observed with the \letg\ spectrometer on board
\chandra. In fact, preliminary analysis of the \agn\ soft X-ray spectra available to us
show features that could, tentatively, be identified as inner-shell 2p - 3d
absorption by M-shell Ca. More details will be presented when the analysis is completed. 

\section{CONCLUSIONS}
\label{sec:concl}

Extensive calculations for inner-shell $n$~= 2 to 3 (mainly 2p~- 3d)
photoexcitation have been carried out for neutral iron (Fe$^{0+}$) and for the
15 ions Fe$^{1+}$ through Fe$^{15+}$. The corresponding absorption forms a Fe
2p~- 3d \uta\ that has been observed in \agn\ spectra between 16~- 17~\AA. It
is shown that despite strong blending within charge-states, the separation
(generally $>$ 0.1~\AA) between the individual ion features enables accurate
diagnostics of the ionization range present in the absorbing medium. Ionic
column densities of up to about 10$^{17}$~cm$^{-2}$ can be relatively easily
diagnosed. Beyond that, the Fe 2p~- 3d \uta\ will be saturated and other
features, such as high-$n$ lines need to be used. Turbulent velocities of
$\le$~700~km/sec can still be marginally resolved. An abbreviated set of
atomic parameters, which can facilitate the spectral modeling of the rather
complex Fe 2p~- 3d \uta, is provided and found in most cases to reproduce
approximately the EW obtained with the full set of atomic
data. Extra caution needs to be applied when using it for low velocities
($\le$~100~km/sec) and ionic column densities of 10$^{17}$~-
10$^{18}$~cm$^{-2}$. Finally, the effects of absorption via 2p~- $n$d
($n~\ge~4$) lines, which are natural to accompany the 2p-3d \uta, should be
observable at ionic column densities of
$\sim$~10$^{17}$~cm$^{-2}$. Conversely, the spectral signature of the
photoionization edges is harder to detect directly, although for ionic column
densities of $\sim$~10$^{17}$~cm$^{-2}$ and higher, their contribution to the
EW is predicted to exceed that of the lines. Finally, it is pointed out that 2p - 3d \uta\ features by lighter elements, such as Ca, Ar, S, and
Si, could also be sought in the soft X-ray and EUV spectra of \agn.

\acknowledgments

  The authors are grateful to the rest of the \rgs\ team for ongoing
  collaboration, which  resulted in the discovery of the Fe 2p~- 3d \uta\ feature and prompted this work. E.B. acknowledges useful discussions with Marcel Klapisch on the physical properties of \uta s. We acknowledge generous support from the National Aeronautics and Space Administration.

\begin{deluxetable}{llcccccrr}
\rotate
\tabletypesize{\small}
\tablecolumns{9}
\tablewidth{0pt}
\tablecaption{Groups of inner-shell $n$~= 2~to 3 absorption lines from the
ground level for Fe$^{0+}$~-~Fe$^{15+}$. \label{tab1}}
\tablehead{
  \colhead{Ion} &
  \colhead{Ground Configuration} &
  \colhead{$J_G$ \tablenotemark{a}} & 
  \colhead{$\lambda_{av}$ (\AA) \tablenotemark{b}} &
  \colhead{$\Sigma f_{ij}$} &
  \colhead{$\Gamma_{eff}$ (10$^{14}$~sec$^{-1}$) \tablenotemark{c}} &
  \colhead{$\Delta \lambda$ (\AA) \tablenotemark{d}} &
  \colhead{LeadingTransition} &
  \colhead{Number of Lines}
}
\startdata

Fe$^{0+}$ & $3s^23p_{1/2}^23p_{3/2}^43d_{3/2}^33d_{5/2}^34s^2$ & 4 & 17.453 &
0.444 & 3.57 & 0.037 & $2p_{3/2} - 3d_{5/2}$ & 59 \\
 & & & 17.142 & 0.109 & 3.77 & 0.025 & $2p_{1/2} - 3d_{3/2}$ & 35 \\
Fe$^{1+}$ & $3s^23p_{1/2}^23p_{3/2}^43d_{3/2}^33d_{5/2}^34s$ & 9/2 & 17.457 &
0.427 & 3.14 & 0.037 & $2p_{3/2} - 3d_{5/2}$ & 64 \\
 & & & 17.148 & 0.105 & 3.50 & 0.027 & $2p_{1/2} - 3d_{3/2}$ & 39\\
Fe$^{2+}$ & $3s^23p_{1/2}^23p_{3/2}^43d_{3/2}^33d_{5/2}^3$ & 4 & 17.420 &
0.459 & 3.45 & 0.041 & $2p_{3/2} - 3d_{5/2}$ & 86 \\
 & & & 17.111 & 0.110 & 2.90 & 0.030 & $2p_{1/2} - 3d_{3/2}$ & 144\\
Fe$^{3+}$ & $3s^23p_{1/2}^23p_{3/2}^43d_{3/2}^23d_{5/2}^3$ & 5/2 & 17.375 &
0.563 & 3.18 & 0.052 & $2p_{3/2} - 3d_{5/2}$ & 62 \\
 & & & 17.065 & 0.149 & 3.24 & 0.036 & $2p_{1/2} - 3d_{3/2}$ & 48 \\
Fe$^{4+}$ & $3s^23p_{1/2}^23p_{3/2}^43d_{3/2}^23d_{5/2}^2$ & 0 & 17.247 &
0.769 & 3.30 & 0.061 & $2p_{3/2} - 3d_{5/2}$ & 17 \\
 & & & 16.943 & 0.191 & 3.11 & 0.054 & $2p_{1/2} - 3d_{3/2}$ & 15 \\
Fe$^{5+}$ & $3s^23p_{1/2}^23p_{3/2}^43d_{3/2}^3$ & 3/2 & 17.179 &
0.797 & 2.97 & 0.056 & $2p_{3/2} - 3d_{5/2}$ & 50\\
 & & & 16.918 & 0.395 & 2.78 & 0.073 &  $2p_{1/2} - 3d_{3/2}$ & 45 \\
Fe$^{6+}$ & $3s^23p_{1/2}^23p_{3/2}^43d_{3/2}^2$ & 2 & 17.033 &
0.988 & 2.41 & 0.079 & $2p_{3/2} - 3d_{5/2}$ & 45 \\
 & & & 16.778 & 0.481 & 2.23 & 0.055 & $2p_{1/2} - 3d_{3/2}$ & 23 \\
Fe$^{7+}$ & $3s^23p_{1/2}^23p_{3/2}^43d_{3/2}$ & 3/2 & 16.887 &
1.061 & 1.66 & 0.082 & $2p_{3/2} - 3d_{5/2}$ & 21 \\
 & & & 16.648 & 0.716 & 1.95 & 0.035 & $2p_{1/2} - 3d_{3/2}$ & 8 \\
Fe$^{8+}$ & $3s^23p_{1/2}^23p_{3/2}^4$ & 0 & 16.775 &
0.688 & 0.96 & 0.020 & $2p_{3/2} - 3d_{5/2}$ & 2 \\
 & & & 16.510 & 1.413 & 1.72 & 0.000 &  $2p_{1/2} - 3d_{3/2}$ & 1 \\
Fe$^{9+}$ & $3s^23p_{1/2}^23p_{3/2}^3$ & 3/2 & 16.614 &
0.598 & 0.74 & 0.074 & $2p_{3/2} - 3d_{5/2}$ & 30 \\
 & & & 16.337 & 1.648 & 1.34 & 0.070 & $2p_{1/2} - 3d_{3/2}$ & 17 \\
 & & & 14.912 & 0.040 & 3.10 & 0.000 & $2s - 3p_{3/2}$ & 1 \\
Fe$^{10+}$ & $3s^23p_{1/2}^23p_{3/2}^2$ & 2 & 16.472 &
0.466 & 0.48 & 0.090 & $2p_{3/2} - 3d_{5/2}$ & 55 \\
 & & & 16.159 & 1.935 & 1.02 & 0.089 & $2p_{3/2} - 3d_{5/2}$ & 60 \\
 & & & 14.754 & 0.084 & 2.66 & 0.008 & $2s - 3p_{3/2}$ & 3 \\
Fe$^{11+}$ & $3s^23p_{1/2}3p_{3/2}^2$ & 3/2 & 16.350 &
0.309 & 0.32 & 0.095 & $2p_{3/2} - 3d_{5/2}$ & 61 \\
 & & & 15.978 & 2.239 & 0.96 & 0.105 & $2p_{3/2} - 3d_{3/2}$ & 90 \\
 & & & 14.591 & 0.133 & 2.24 & 0.014 & $2s - 3p_{1/2}$ & 8 \\
Fe$^{12+}$ & $3s^23p_{1/2}^2$ & 0 & 16.209 &
0.275 & 0.75 & 0.118 & $2p_{3/2} - 3d_{3/2}$ & 21 \\
 & & & 15.787 & 2.418 & 1.45 & 0.093 & $2p_{3/2} - 3d_{5/2}$ & 27 \\
 & & & 14.446 & 0.178 & 2.00 & 0.029 &  $2s - 3p_{3/2}$ & 4 \\
Fe$^{13+}$ & $3s^23p_{1/2}$ & 1/2 & 16.033 &
0.224 & 0.78 & 0.149 & $2p_{3/2} - 3d_{3/2}$ & 34 \\
 & & & 15.594 & 2.612 & 0.84 & 0.103 & $2p_{1/2} - 3d_{3/2}$ & 26 \\
 & & & 14.288 & 0.227 & 1.33 & 0.028 & $2s - 3p_{3/2}$ & 6 \\
Fe$^{14+}$ & $3s^2$ & 0 & 15.622 &
0.932 & 0.46 & 0.137 &  $2p_{3/2} - 3d_{5/2}$ & 15 \\
 & & & 15.334 & 2.063 & 1.09 & 0.042 & $2p_{1/2} - 3d_{3/2}$ & 2 \\
 & & & 14.124 & 0.282 & 0.51 & 0.019 & $2s - 3p_{3/2}$ & 2 \\
Fe$^{15+}$ & $3s$ & 1/2 & 17.296 &
0.087 & 0.03 & 0.141 & $2p_{3/2} - 3s$ & 2 \\
 & & & 15.514 & 0.505 & 0.35 & 0.081 & $2p_{3/2} - 3d_{5/2}$ & 18 \\
 & & & 15.191 & 2.536 & 0.45 & 0.072 & $2p_{1/2} - 3d_{3/2}$ & 7 \\
 & & & 13.966 & 0.303 & 0.82 & 0.068 & $2s - 3p_{3/2}$ & 6 \\

\enddata
\tablenotetext{a}{ total angular momentum quantum number of the ground level.}
\tablenotetext{b}{ defined in Eq. (5).}
\tablenotetext{c}{ defined in Eq. (6).}
\tablenotetext{d}{ defined in Eq. (7).}
\end{deluxetable}

\clearpage

\begin{figure}
  \plotone{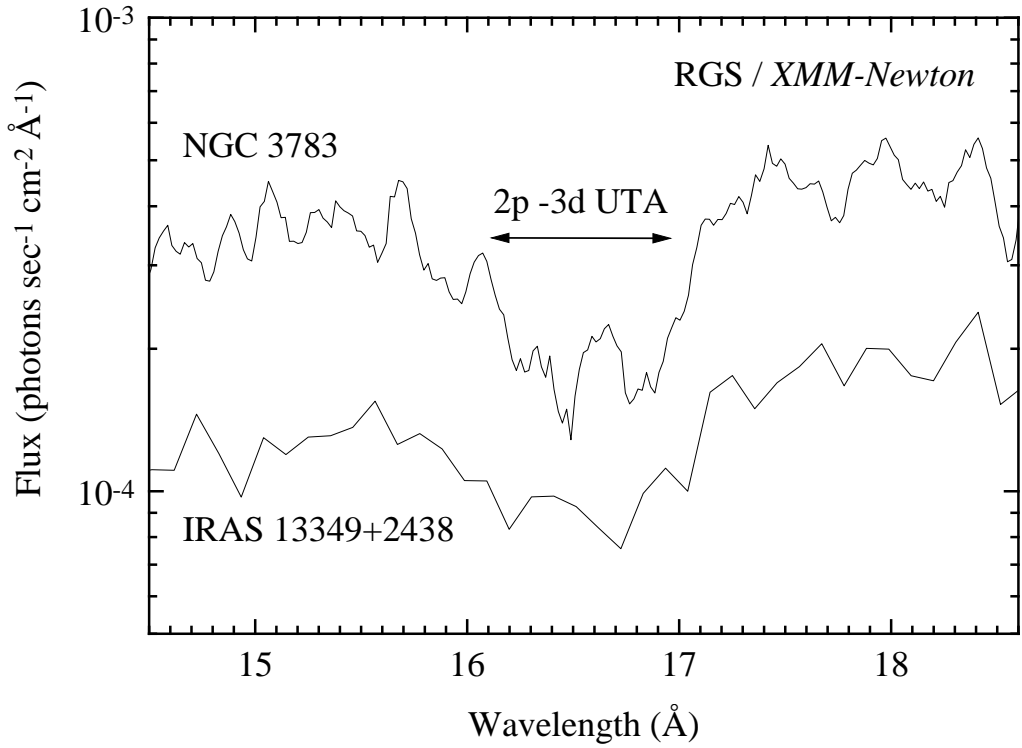}
  \caption{The RGS spectra of the Seyfert~1 NGC 3783 and the luminous Quasar IRAS 13349+2438 featuring a broad trough between 16 - 17~\AA, due to numerous inner-shell $n$= 2 to 3 (mainly 2p~- 3d) absorption lines of Fe M-shell ions blended into a \uta.}  \label{f1}
\end{figure}

\begin{figure}
  \plotone{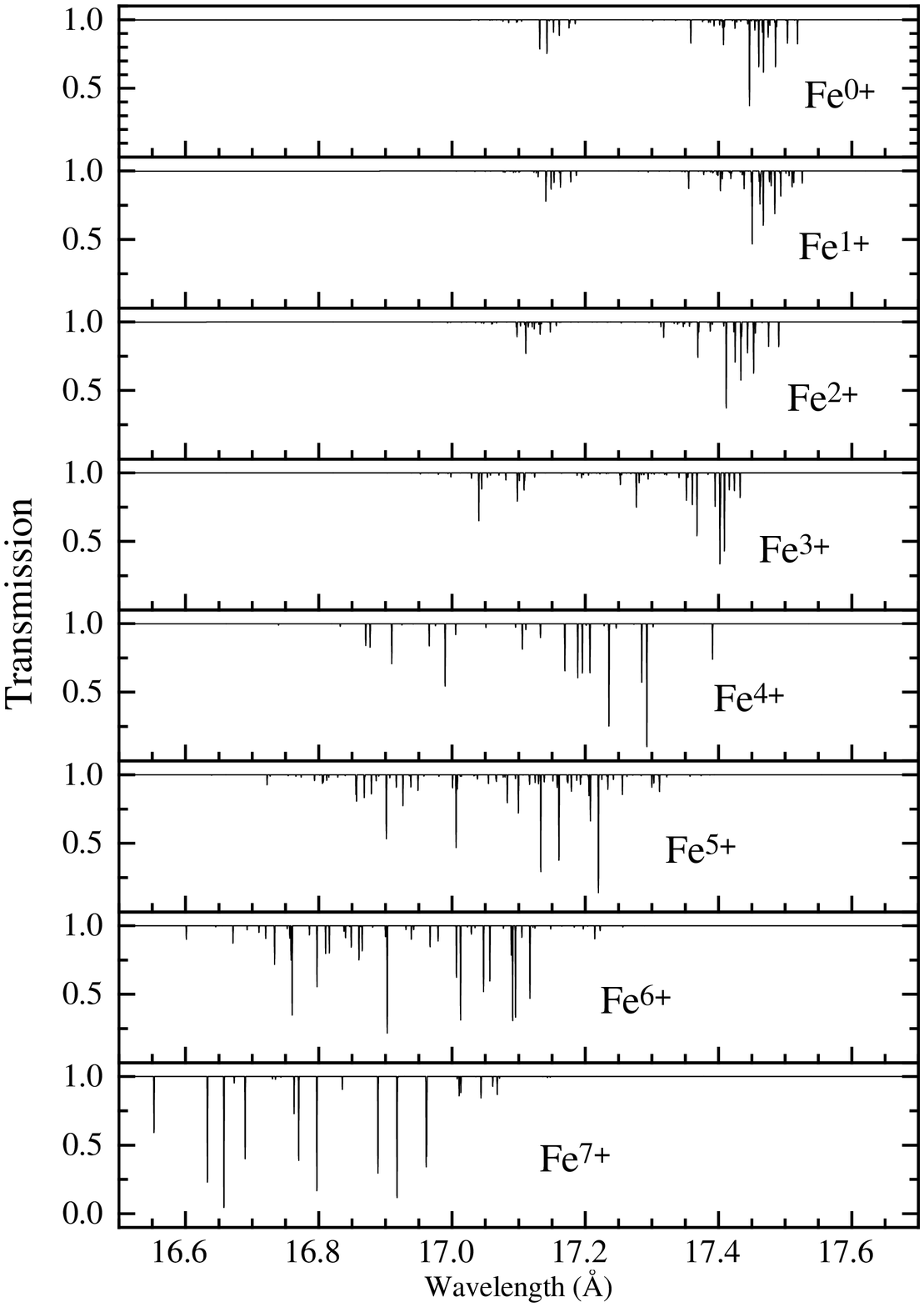}
  \caption{Spectra of inner-shell $n$~= 2 to 3 absorption lines for Fe$^{0+}$
  through Fe$^{7+}$. A temperature of $kT_{ion}$~= 10~eV and no turbulent
  velocity are assumed for these plots, in order to minimize line blending and
  emphasize the complex atomic structure and rich line spectra. The ionic 
  column density used is 2 $\times10^{15}$ cm$^{-2}$.} \label{f2}
\end{figure}

\begin{figure}
  \plotone{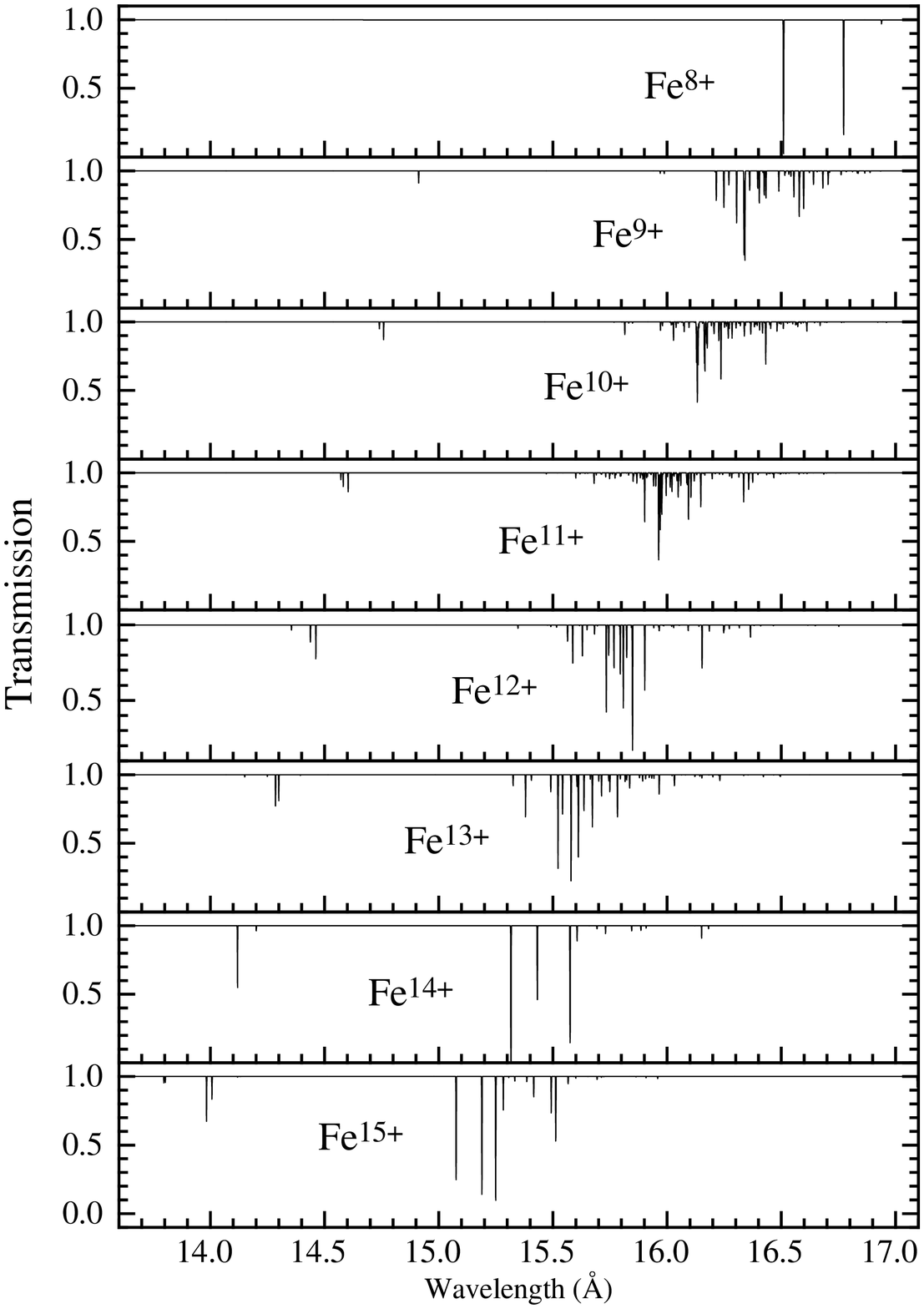}
  \caption{Spectra of inner-shell $n$~= 2 to 3 absorption lines for Fe$^{8+}$
  through Fe$^{15+}$. A temperature of $kT_{ion}$~= 10~eV and no turbulent
  velocity are assumed for these plots, in order to minimize line blending and
  emphasize the complex atomic structure and rich line spectra. The ionic 
  column density used is 2 $\times 10^{15}$ cm$^{-2}$.} \label{f3}
\end{figure}

\begin{figure}
  \plotone{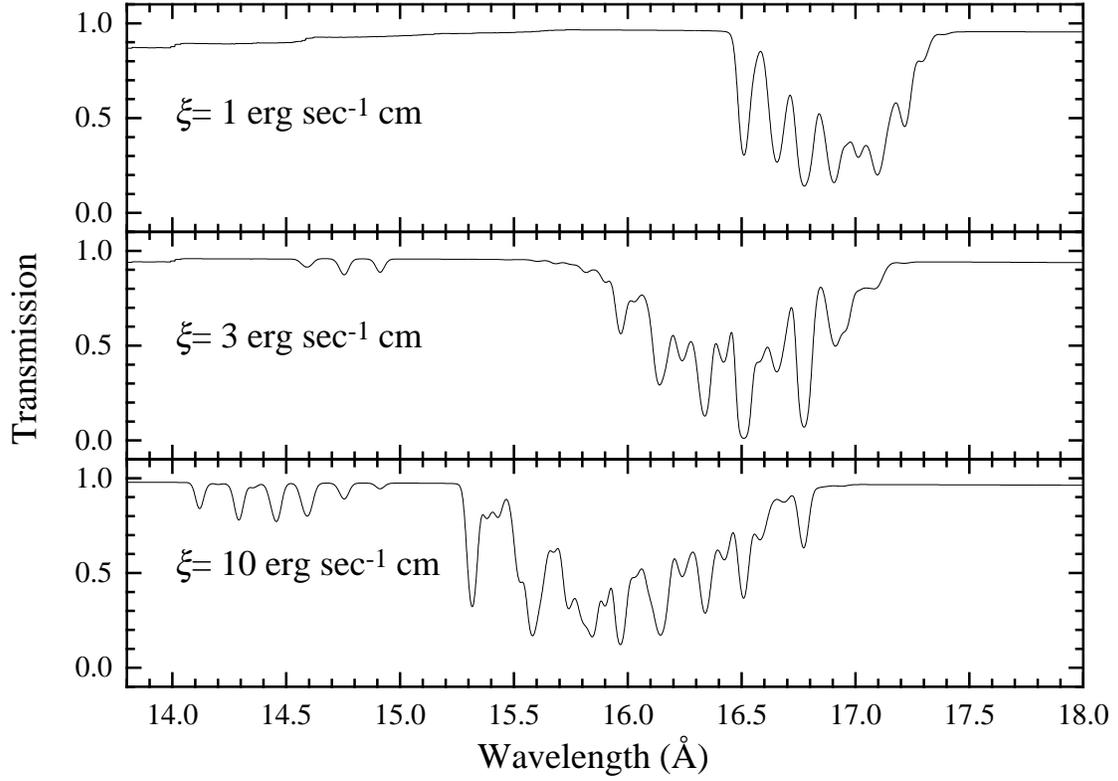}
  \caption{Multi-ion spectra featuring numerous inner-shell $n$~= 2 to 3
  (mainly 2p~- 3d) absorption lines of Fe M-shell ions blended into a
  \uta. Calculations are carried out for single ionization-parameter zones
  ($\xi = L / n_er^2$). $N_{Fe}$~= 2$\times10^{17}$~cm$^{-2}$ and $v_{turb}$~=
  500~km/sec are assumed for the three plots.} \label{f4}
\end{figure}

\begin{figure}
  \plotone{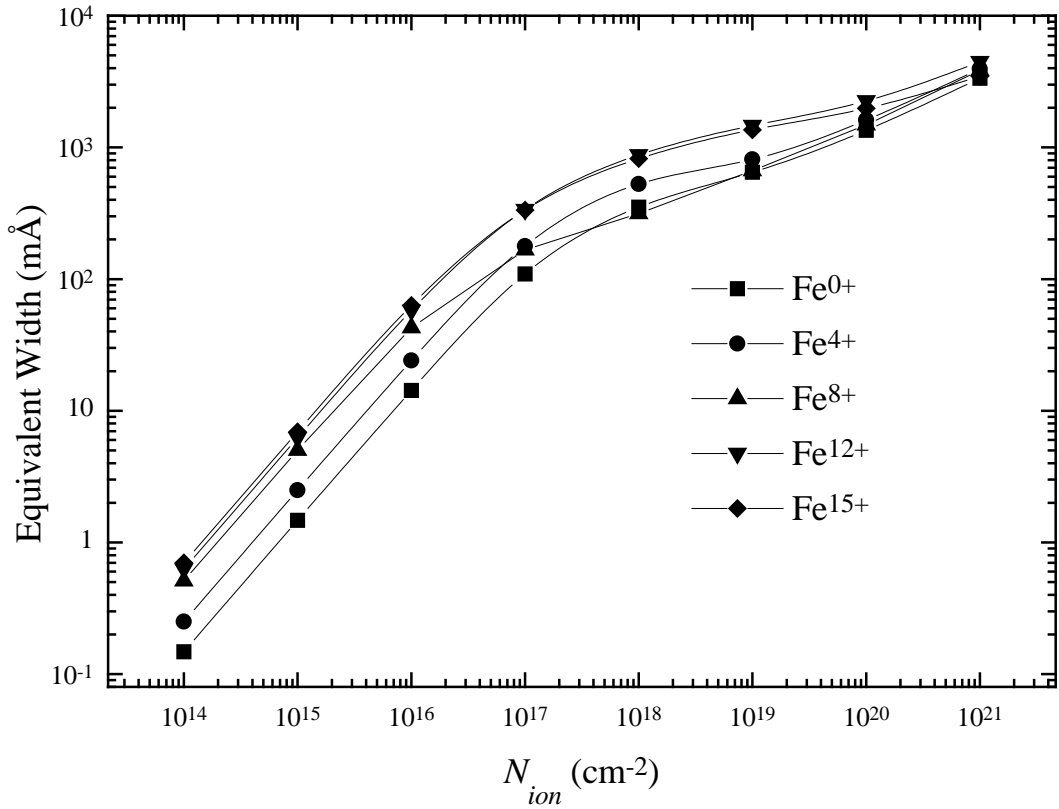}
  \caption{Curves of growth for five selected Fe M-shell ions due to the
  inner-shell $n$ = 2 to 3 (mainly 2p~- 3d) absorption lines. EW is calculated in the wavelength region between 10 and 20~\AA. $v_{turb}$ is taken to be 500~km/sec.} \label{f5}
\end{figure}

\begin{figure}
  \plotone{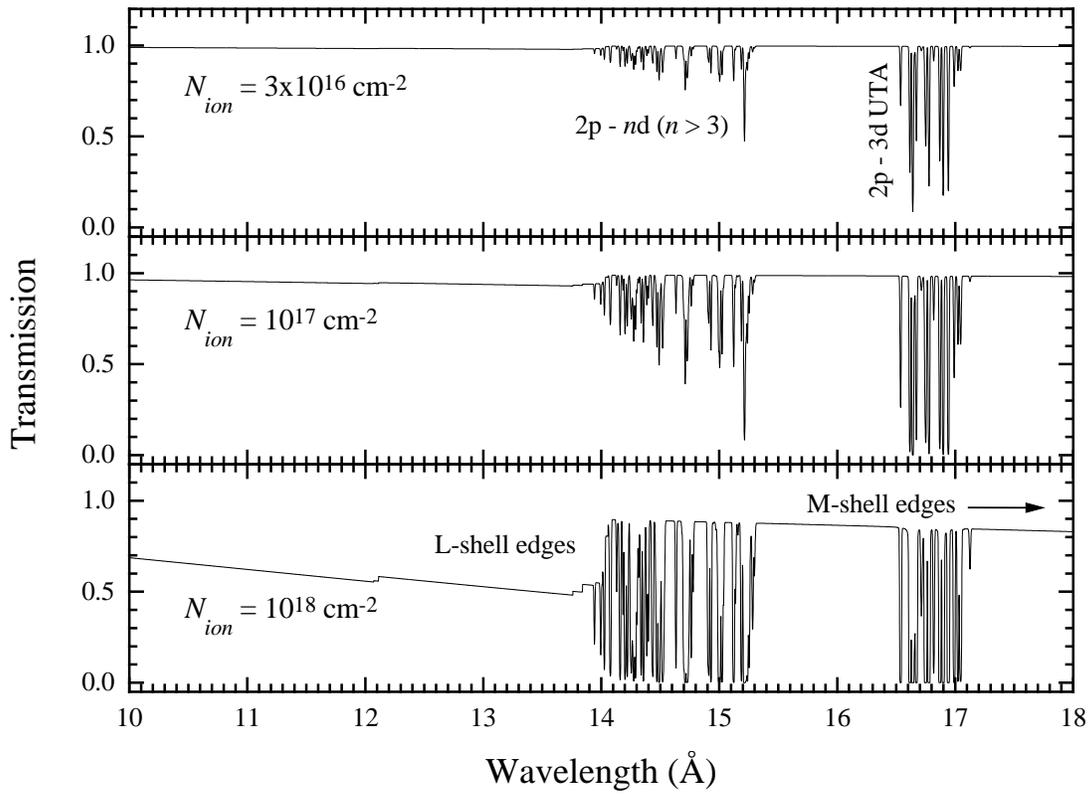}
  \caption{Spectra of Fe$^{7+}$ at three different ionic column densities. Calculations include all of the inner-shell 2p~- $n$d ($n$~= 3 to 9) absorption lines, as well as the L-shell and M-shell photoionization edges. The turbulent velocity is fixed at 100~km/sec.} \label{f6}
\end{figure}

\begin{figure}
  \plotone{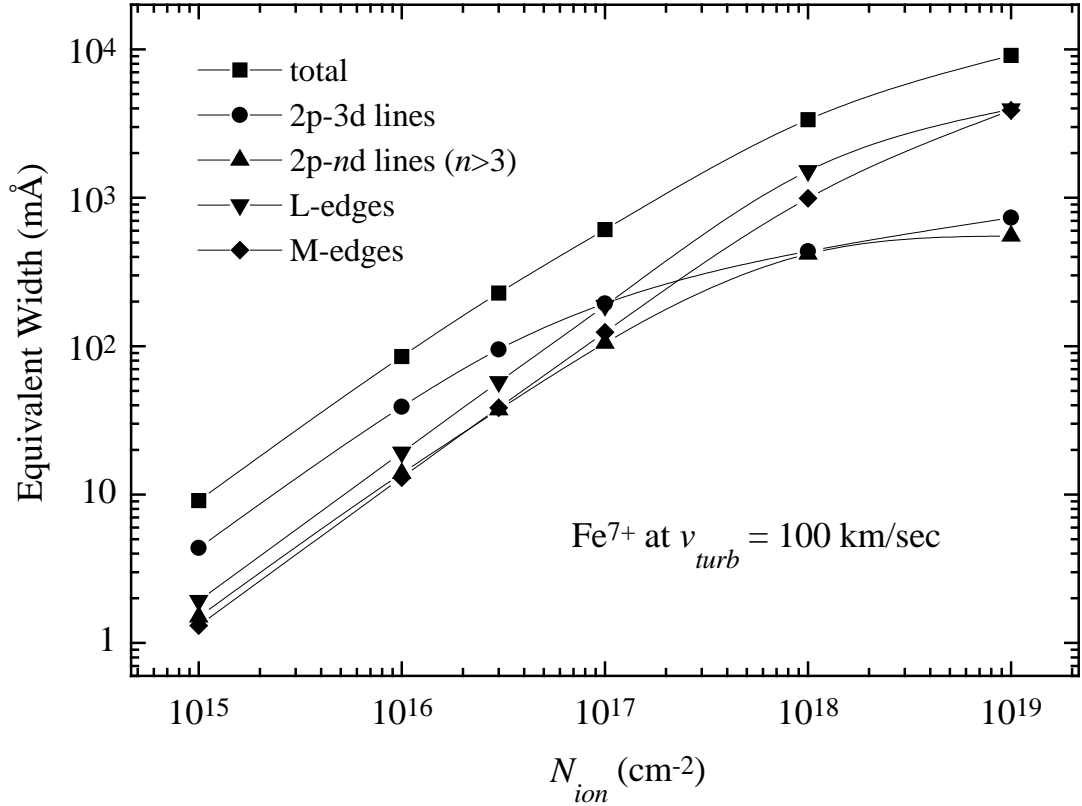}
  \caption{Curves of growth for Fe$^{7+}$. Separate curves are given for the
  EW contributions of inner-shell 2p~- 3d absorption lines, inner-shell 2p~-
  $n$d ($n$~= 4 - 9) absorption lines, and for the L-shell and M-shell
  photoionization edges. EW is calculated in the wavelength region between 10
  and 20~\AA. $v_{turb}$ is taken to be 100~km/sec.} \label{f7}
\end{figure}

\end{document}